\documentclass{an}
\usepackage{graphicx}
\usepackage{times}
\Volume{01}              % issue in current year, format {NN}
\Year{2002}              % format: {YYYY}
\Month{10}               % format: {M} or {MM} (number), month of print version
\Pagespan{001}{002}      % format: {start page}{last page}

\begin{document}
\lhead[\thepage]{A.N. Author: Title}
\rhead[Astron. Nachr./AN~{\bf XXX} (200X) X]{\thepage}
\headnote{Astron. Nachr./AN {\bf 32X} (200X) X, XXX--XXX}

\title{Astrometric confirmation of a wide low-mass companion to the planet host star HD\,89744}

\author{Markus~Mugrauer\inst{1}, R.~Neuh\"auser\inst{1},
T.~Mazeh\inst{2}, E.~Guenther\inst{3} \and M.~Fern\'andez\inst{4}}
\institute{Astrophysikalisches Institut und Universit\"ats-Sternwarte
Jena, Schillerg\"asschen 2-3, 07745 Jena, Germany\and Tel Aviv
University, Tel Aviv 69978, Israel \and Th\"uringer Landessternwarte,
Sternwarte 5, 07778 Tautenburg, Germany\and Instituto de Astrof\'isica
de Andaluc\'ia, CSIC, Apdo.  Correos 3004, 18080 Granada, Spain}
\date{Received {date will be inserted by the editor}; accepted {date
will be inserted by the editor}}

\abstract{In our ongoing deep infrared imaging search for faint wide secondaries of
planet-candidate host stars we have confirmed astrometrically the companionship of a low-mass
object to be co-moving with HD\,89744, a companion-candidate suggested already by
Wilson~et~al.~(2000). The derivation of the common proper motion of HD\,89744 and its companion,
which are separated by 62.996$\pm$0.035\,arcsec, is based on about 5\,year epoch difference between
2MASS and our own UKIRT/UFTI images. The companion effective temperature is about 2200\,K and its
mass is in the range between  0.072 and 0.081\,$M_{\sun}$, depending on the evolutionary model.
Therefore, HD\,89744\,B is either a very low mass stellar or a heavy brown dwarf companion.
}
\correspondence{markus@astro.uni-jena.de}

\maketitle

\section{Introduction}
Up to now precise radial velocity (rad-vel) searches have studied approximately 2000 sun-like
stars and found so far more than one hundred extra-solar planets (Lineweaver\&Grether~2003), some
of which are found in stellar binary systems (e.g. Naef~et~al.~2003). The planets found in the
binaries are most intriguing, because they provide the possibility to study the effect of stellar
multiplicity on the planet formation and the long-time stability and evolution of planetary orbits
(Zucker \& Mazeh~2002).

Companions close to the planet-hosting stars (closer than 100\,AU) can be detected either
indirectly by rad-vel measurements (e.g. Korzennik~et~al.~2000) or directly by high resolution
speckle (e.g. Neuh\"auser et al. 2000) or adaptive optics (AO) imaging (e.g.
Patience~et~al.~2002). Wide companions, with separations larger than 100\,AU, lie outside the
field of view (FOV) of AO systems, and can not be detected by rad-vel variation, because of their
long orbital periods. Such companions are, however, reachable by wide field imagers but so far the
whole sample of extra-solar planetary systems has not been surveyed homogeneously for wide
companions with sensitive IR cameras. We therefore initiated a systematic search for wide faint
companions to all stars known to have rad-vel planets, using relatively large IR images with up to
150\,arcsecs FOV. The intrinsic faintness of low-mass companions can be confirmed by spectroscopy,
and their companionship can be established by detecting proper motion equal to that of the
planet-hosting star, which usually has large proper motion ($\mu$\,$\sim$\,200\,mas/yr) well known
due to precise measurements of the European astrometry satellite \textsl{HIPPARCOS}. The first
result of our project is presented here.

A faint wide candidate for a companion of one of the stars orbited by a planet was already
suggested by Wilson~et~al.~(2001), who searched the 2MASS database for red companions to nearby
stars. They found a very red object close to HD\,89744, which is now known to have a planet with a
minimum mass ($msin(i)$) of 7.2\,M$_{Jup}$ a period of 256 days with a semi-major axis ($a$) of
0.88\,AU on a very eccentric orbit, with $e$\,=\,0.7 (Korzennik~et~al.~2000). Wilson~et~al.\
obtained a spectra of the faint object and derived a spectral type of L0V. Although the 2MASS
photometry is consistent with a L0V object at the distance of HD\,89744, the companionship could
not be established beyond any doubt  without further astrometry that will show that the two
objects share the same proper motion. This paper presents astrometric observations that show that
the faint companion of Wilson~et~al. does indeed move with the same proper motion as HD\,89744,
establishing the physical association of the two stars.

\section{Observations and Data Reduction}
All observations of our program were done in the H-band (1.6\,$\mu$m), because late-type stars and
substellar objects are much brighter in near IR than in the visible bands, and IR detectors
available today offer sufficient sensitivity. Therefore the contrast between the bright primary
(hot) and a low-mass close-in companions (cool) is less prominent in the near infrared than in the
visible, rendering the companion detection in the IR possible. Observations were made at the United
Kingdom Infrared Telescope (UKIRT) with the IR camera \textsl{UKIRT\,\,Fast-Track\,\,Imager}
(UFTI), a $1024\times1024$ HgTeCd-detector with a pixel resolution of 90.85$\pm$0.2\,mas/pixel,
i.e., 93\,arcsecs FOV. UFTI can detect an $H$\,$\sim$\,20\,mag star in 10\,min total integration
time (see sect.\,5).

We used sky-flatfielding to correct the different pixel sensitivity and the standard IR imaging
jitter technique for background subtraction. The flatfield image was composed of several images of
a faint (not saturating) standard star. The integration time must be long enough ($>$\,10\,s) to
get a good sky background signal which can be used as skyflat. The standard star was observed on
different chip positions (jitter-technique). A point spread function (PSF) was fitted to this star
and it was subtracted from the individual images. The PSF subtracted images contained only
information of the chip illumination and were median combined to create the flatfield image. Both
flatfielding and background subtraction were carried out using ESO package \textsl{ECLIPSE}, with
which we also combined all the reduced frames into the resulting image. Due to the brightness of
the planet-hosting star we had to use short integration times, in order to minimize saturation
effects. With a 4s integration time (the shortest possible UFTI integration time) saturation
occurred only in the central region of the stellar PSF. The raw images were nearly background
limited (6s with UFTI at UKIRT), which gave us a good sensitivity for faint wide companions.
Observations were made with a 8-point-jitter cycle on the IR detector. To reduce the overhead, on
each jitter-position six short 4\,s integrations were added, hence 24\,s total integration time per
jitter-position. We repeated the 8-point-jitter cycle three times hence 3x(8x24\,s)=576\,s total
integration time per target.

\section{Astrometry}

We have obtained two H-band images of HD\,89744 with UFTI in June 2002 and June 2003. Several
companion-candidates are detected in the images, all with S/N$>>$10. Figure~1 shows one of these
images, together with a previous IR observation of HD\,89744 which had been taken by the Two Micron
All Sky Survey (2MASS) in April 1998. Three objects are listed in the 2MASS point source catalog,
namely the companion-candidates~1 and 6, together with HD\,89744 itself.
\begin{figure}[ht]
\resizebox{\hsize}{!} {\includegraphics[]{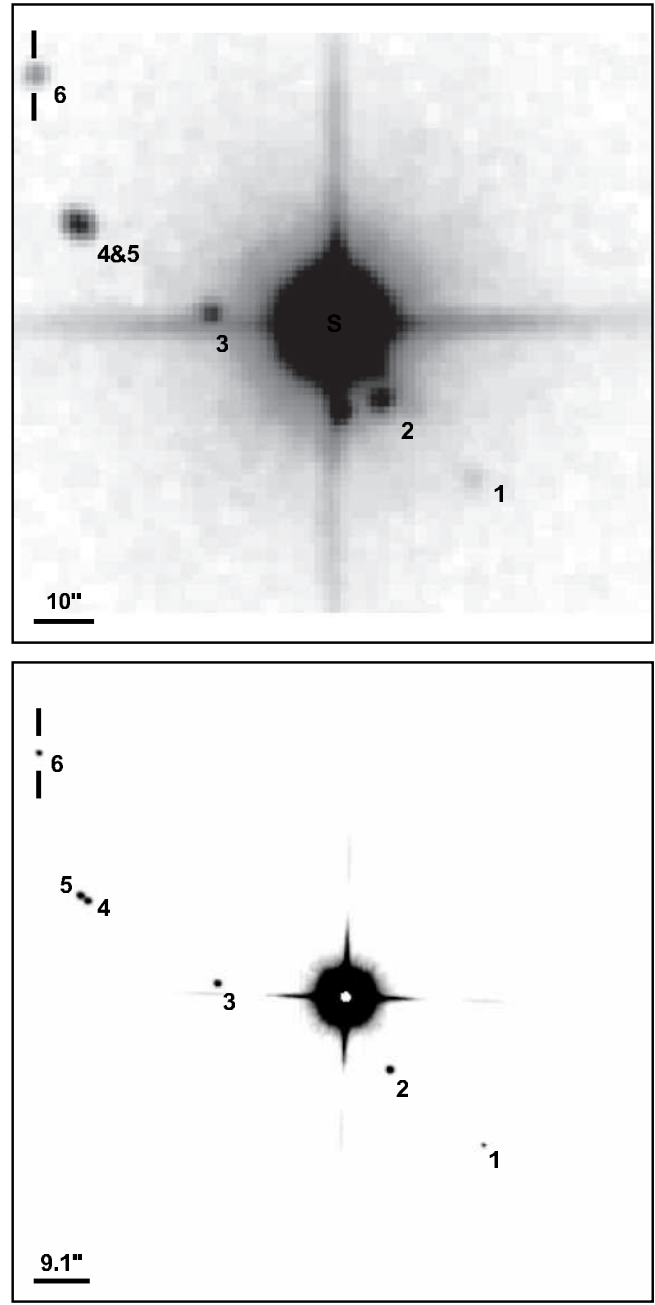}} \caption{H-band images of HD\,89744. Top
--- 2MASS image (1998), bottom --- UFTI image (2003) . North is up and east to the left. The bright rad-vel
planet host star is located in the center of the image.  Several companion-candidates are detected.
The co-moving companion HD\,89744\,B can be found in the upper left corner of the images
(candidate~6).} \label{picture}
\end{figure}

\begin{table}[ht]
\caption{Table of observations} \label{tablabel1}
\begin{tabular}{ccccc}
\hline
instrument & epoch & pixelscale & FWHM\\
           &       & (arcsec)   & (arcsec)\\
\hline UFTI/UKIRT & 05/06/02 & 0.09085 & 0.64\\
\hline UFTI/UKIRT & 10/06/03 & 0.09085 & 0.55\\
\hline\hline
\end{tabular}
\end{table}

By using the separations in the 2MASS image as a starting value and using \textsl{HIPPARCOS} data
for the parallactic and proper motion of HD\,89744, one can determine the expected separations
between the candidates and HD\,89744 for our two epochs, assuming the candidates are non-moving
background stars. Physical companions should have same proper motion as the primary, therefore
their position relative to the primary should be constant. We can neglect orbital motion here,
because for wide companions (a$>$100\,AU) it is much smaller than the proper and parallactic
motion.

\begin{figure}[ht]
\resizebox{\hsize}{!} {\includegraphics[]{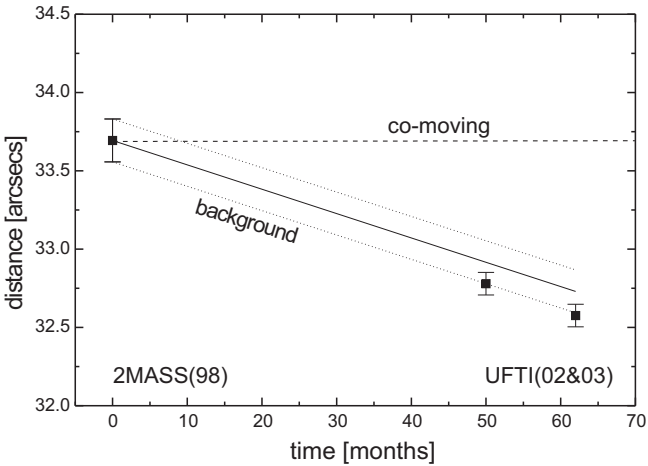}} \caption{The measured separation
between HD\,89744 and the companion-candidate~1 for 2MASS (1998) and UFTI (2002\,\&\,2003) images.
The separation (black data points) becomes smaller exactly as it is predicted by the parallactic
and proper motion of HD\,89744 (see full line), hence this candidate is a non-moving background
star.} \label{astrometry1}
\end{figure}

\begin{figure}[ht]
\resizebox{\hsize}{!} {\includegraphics[]{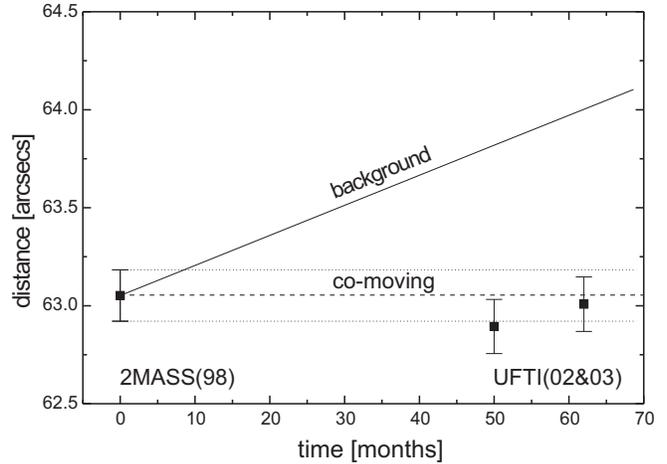}} \caption{The separation between
HD\,89744 and the companion-candidate~6. The measured separations (black data points) are constant
for the given range of time. Therefore, candidate~6 is co-moving with HD\,89744.}
\label{astrometry2}
\end{figure}

The expected positions of candidates~1 and 6, both as co-moving and as background objects, are
shown in Fig.\,\ref{astrometry1}\,\&\,\ref{astrometry2}, together with the actual separation of the
two objects. The distance between HD\,89744 and candidate~1 became smaller between 1998 and 2003,
exactly as predicted by the \textsl{HIPPARCOS} data of the primary. Hence companion-candidate~1 is
a non-moving background star. On the other hand, the distance between HD\,89744 and candidate~6
remained constant within the astrometric precision in all three epochs. We therefore conclude that
candidate~6 is a co-moving companion to the planetary system. Consequently, candidate~6 must be
physically associated with HD\,89744, because the chance to find a fore-/background object with
exactly the same proper motion is negligible. Candidate~6 has common proper motion (CPM) with
HD\,89744 and will be denoted as HD\,89744\,B in the following.

The two UFTI observations taken with only one year apart, the data from the 2MASS point source
catalogue and the 2MASS image itself can be used to check whether any of the remaining detected
candidates~2, 3, 4 or 5 are co-moving with the planetary system. In Table\,\ref{tablabel2} we have
listed the results of the astrometry test for these candidates. Candidates~2 and 3 do not move
together with HD\,89744. While the evidence for candidate~2 is quite strong ($\sim$\,7\,$\sigma$
confidence level), the evidence for candidate~3 is not that conclusive (only $\sim$\,2\,$\sigma$
level) by using only our UFTI images. Candidate~3 is not listed in the 2MASS point source catalogue
but is well separated from the primary in the 2MASS image (see Fig.\,\ref{picture}). We measure the
positions of both objects in the 2MASS image and get a separation of 19.925$\pm$0.148\,arcsecs for
the epoch 1998. Hence with the 2MASS image and our UFTI images we can conclude that candidate~3 is
also a none-moving background star with a confidence level of $\sim$5$\sigma$. The
Table\,\ref{tablabel2} shows that candidates~4 and 5 do not move relative to candidate~3.

\begin{table}[ht]
\caption{Astrometry for the two UFTI observations. The separation between all detected objects is
listed. The motion of the star is clearly detectable (see S-2 and S-3 for separation between the
primary star (S) and the companion-candidates~2 and 3, respectively). The separations 3-4, 3-5 and
4-5 are constant for both epochs.} \label{tablabel2}
\begin{tabular}{ccc}\hline
objects & separation 2002 & separation 2003\\
        & (arcsec)          & (arcsec)\\
\hline
S-2 & 13.913$\pm$0.031 & 13.688$\pm$0.031 \\
\hline
S-3 & 20.620$\pm$0.045 & 20.720$\pm$0.046 \\
\hline
3-4 & 24.771$\pm$0.055 & 24.747$\pm$0.055 \\
\hline
3-5 & 26.186$\pm$0.058 & 26.176$\pm$0.058 \\
\hline
4-5 & \,\,1.418$\pm$0.012 & \,\,1.432$\pm$0.012 \\
\hline\hline
\end{tabular}
\end{table}

\section{The Nature of the two Members of the Common Proper Motion Pair}
HD\,89744 is a F7V (B-V=0.531) star located at a distance of 39$\pm$1\,pc (Perryman 1997). Its mass
is 1.4$\pm$0.2 (see Allende Prieto \& Lambert 1999; Ng \& Bertelli 1998; Feltzing 2001; Chen \&
Zhao 2002, Laws 2003; Santos 2004; Fuhrmann 2004). The stellar age ranges from 2 to 3\,Gyrs (see
Marsakov 1995; Feltzing 2001; Ibukiyama 2002; Chen 2002; Laws 2003). With $[Fe/H]$\,=\,0.18 (Taylor
2003) HD\,89744 is a bit more metal rich than the sun, a frequent feature for stars hosting rad-vel
planets (Udry\&Mayor 2002).

2MASS photometry of HD\,89744\,B in the $J$,$H$ and $K$ bands and the distance to HD\,89774
yielded absolute near IR magnitudes for the faint companion, all given in Table\,\ref{tablabel3}.
Our relative aperture photometry of our UFTI images is consistent with the H band mag of 2MASS,
within the photometry precision (see Table\,\ref{tablabel3}).

\begin{table}[ht]
\caption{2MASS photometry for HD\,89744\,B and the calculated absolute magnitude in $J$,$H$ and
$K$. In $H$ we show also the UFTI photometry results.} \label{tablabel3}
\begin{tabular}{ccc}\hline
band & app. magnitude & abs. magnitude\\
     & (mag)            & (mag)\\
\hline
$J$ & 14.901$\pm$0.037 & 11.946$\pm$0.071 \\
\hline
$H$ & 14.022$\pm$0.033 & 11.067$\pm$0.071\\
$H_{UFTI}$ & 14.070$\pm$0.052 & 11.115$\pm$0.080\\
\hline
$K$ & 13.608$\pm$0.039 &10.653$\pm$0.072 \\
\hline\hline
\end{tabular}
\end{table}

Several evolutionary models for low-mass stars, brown dwarfs and planets are available today that
can determine the faint object mass and effective temperature from the obtained absolute
photometry. Table\,\ref{tablabel4} shows the results of three models from Baraffe (1998 and 2003)
and Chabrier~et~al.~(2000), namely BCAH98, DUSTY00 and COND03. For BCAH98 we used  the model with
mixing length of $\alpha$\,=\,1.0, He abundance $Y$\,=\,0.275 and solar metallicity $[M/H]$\,=\,0,
because it is the only set available for low-mass objects down to 0.072\,$M_{\sun}$.
Table\,\ref{tablabel4} shows results of all models for an age of 1 and 5\,Gyrs. For an age of
1\,Gyr HD89744\,B has an averaged mass of 0.076\,$M_{\sun}$ and an effective temperature of
2200\,K, whereas for 5\,Gyrs one gets m\,=\,0.079\,$M_{\sun}$ and $T_\mathrm{eff}$\,=\,2200\,K
(average of $T_\mathrm{eff}$ always rounded to nearest 100\,K). Therefore we can conclude that the
co-moving companion HD\,89744\,B is either a heavy brown dwarf or a very low-mass stellar
companion.

\begin{table}[htb]
\caption{The mass and effective temperature of HD\,89744\,B. Models used were BCAH98, DUST00 and
COND03 for 1 and 5\,Gyrs. Values of T$_\mathrm{eff}$ always rounded to nearest 100\,K.}
\label{tablabel4}
\begin{tabular}{ccc}\hline
model & mass        & T$_\mathrm{eff}$\\
      & (M$_{\sun}$)& (K)\\
\hline& *** age 1Gyr *** &\\
\hline
BCAH98 & 0.074$\pm$0.002 & 2200\\
\,\,\,DUSTY00 & 0.077$\pm$0.002 & 2200\\
COND03 & 0.076$\pm$0.003 & 2300\\
\hline
& *** age 5Gyr *** &\\
\hline
BCAH98 & 0.078$\pm$0.001 & 2200\\
\,\,\,DUSTY00 & 0.080$\pm$0.001 & 2200\\
COND03 & 0.078$\pm$0.003 & 2200\\
\hline\hline
\end{tabular}
\end{table}

\section{Discussion and Conclusion}

\begin{figure}[ht]
\resizebox{\hsize}{!} {\includegraphics[]{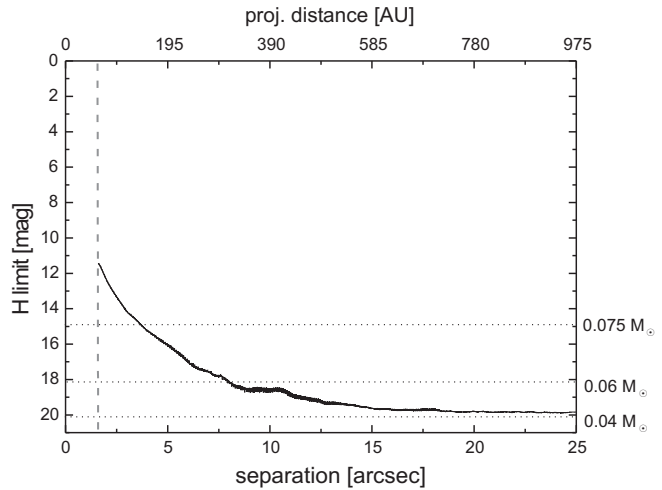}} \caption{The UFTI detection limit (S/N=3)
for a range of separations. The averaged FWHM of all detected objects is $\sim$0.6\,arcsecs. Inside
a radius of 62\,AU ($\sim$1.6\,arcsec) around the bright rad-vel planet host star saturation
occurs. This part of the detector cannot be used for companion detection and is excluded in the
plot. The detection of all stellar companions down to 144\,AU ($\sim$\,3.7\,arcsec) is possible. We
assumed a conservative system age of 5\,Gyrs and used Baraffe COND03 models for the magnitude-mass
conversion.} \label{dynamic}
\end{figure}

By measuring the spatial noise level in the two UFTI images we can determine the UFTI detection
limit (see Fig.\,\ref{dynamic}). $H$\,$\sim$\,20\,mag is reached in the background noise dominated
region, which allows us to find even substellar companions down to 0.040\,$M_{\sun}$ assuming a
conservative star age of 5\,Gyrs and use Baraffe COND03 models. Closer to the star the detection
limit is worse due to the increasing photon noise of the nearby bright star.  With the given UFTI
detection limit shown in Fig.\,\ref{dynamic} and the results from the UFTI astrometry in sec.\,3
we can conclude that there is no further stellar companion around HD\,89744\,A from 144\,AU
(3.7\,arcsecs) up to 1891\,AU (48.5\,arcsecs), which is the maximal FOV completely covered by both
UFTI images. Further wider companions can be detected up to 2573\,AU ($\sim$\,66\,arcsec) but only
64\% of this larger FOV is covered by the UFTI detector.

Can we detect the orbital motion of HD\,89744\,A and B?  The angular separation between the two of
them is 62.996$\pm$0.035\,arcsec, i.e., 2456$\pm$67\,AU projected separation. For $a$\,=\,2500\,AU
and a total mass of about 1.4\,$M_{\sun}$ the orbital period is approximately 10$^{5}$\,years,
hence too long to detect any astrometric or rad-vel orbital motion.

Finally we can estimate limits of long-term stability criteria for further undiscovered companions
in the HD\,89744 system. Holman \& Wiegert (1999) found a certain critical semi-major axis $a_{c}$
for which circular orbits of test particles around the primary will be disrupted, depending on the
eccentricity of the wide companion. This critical semi-major axis can be further reduced if the
test particles themselves are on eccentric orbits. For a mass ratio
$\mu$\,=\,$m_{S}/(m_{P}+m_{S})$\,=\,0.1 and a presumed circular orbit of HD\,89744\,B
($a$$\sim$2500\,AU) we get $a_{c}$$\sim$\,1100\,AU and $a_{c}$$\sim$\,100\,AU for a very eccentric
orbit with $e$\,=\,0.8. However the mass ratio of the HD\,89744 system
($m_{P}$\,$\sim$\,1.4\,$M_{\sun}$ and $m_{S}$\,$\sim$\,0.077\,$M_{\sun}$ $\Rightarrow$
$\mu$\,=\,0.052) is smaller, i.e., $a_{c}$ is even larger as calculated above. With the Holman \&
Wiegert stability criteria we can conclude that further companions could reside within a 100\,AU
orbit even for very eccentric orbits of HD\,89744\,B. Those objects are detectable by AO search
programs which are on the way. But so far no further low-mass companion is reported in the
HD\,89744\,A\&B system.

A few planet-hosting stars have faint companions like HD\,89774. E.g. HD\,114762, $\upsilon$\,And
and 55\,CnC have common proper motion companions with projected separations of 131$\pm$8\,AU,
747$\pm$8\,AU and 1060$\pm$13\,AU respectively (separations determined with our images). The set
of planets in wide binaries is still not complete, and therefore no comparison with planets in
close binaries and around single stars is possible. Such a comparison might be possible when our
systematic search would be complete.

\acknowledgements {We would like to thank the technical staff of UKIRT for all their help and
assistance in carrying out the observations. The United Kingdom Infrared Telescope (UKIRT) is
operated by the Joint Astronomy Centre on behalf of the U.K. Particle Physics and Astronomy
Research Council. This publication made use of data products from the Two Micron All Sky Survey,
which is a joint project of the University of Massachusetts and the Infrared Processing and
Analysis Center/California Institute of Technology, funded by the National Aeronautics and Space
Administration and the National Science Foundation. We have used the SIMBAD database, operated at
CDS, Strasbourg, France. Furthermore we would like to thank A.~Seifahrt, A.~Szameit and C.~Broeg
who have carried out some of the observations of this project. This work was partly supported by
the Israel Science Foundation (grant no.  233/03)}


\begin{thebibliography}{}
\bibitem{allende}Allende Prieto, C., Lambert, D.L.: 1999, A\&A 352, 555
\bibitem{baraffe1}Baraffe, I., Chabrier, G., Allard, F.: 1998, A\&A 337, 403
\bibitem{baraffe2}Baraffe, I., Chabrier, G., Barman, T.S.: 2003, A\&A 402, 701
\bibitem{chabrier}Chabrier, G., Baraffe, I., Allard, F.: 2000, ApJ 542, 464
\bibitem{chen}Chen, Yu-Qin, Zhao, Gang: 2002, ChJAA 2, 151
\bibitem{feltzing}Feltzing, S., Holmberg, J., Hurley, J.R.: 2001, A\&A 377, 911
\bibitem{fuhrmann}Fuhrmann, K.: 2004, AN 325, 3
\bibitem{holman}Holman, M.J., Wiegert, P.A.: 1999, AJ 117, 621
\bibitem{ibukiyama}Ibukiyama, A., Arimoto, N.: 2002, A\&A 394, 927
\bibitem{korzennik}Korzennike, S.G., Brown, T.M., Fischer, D.A.: 2000, ApJ 533, 147
\bibitem{laws}Laws, C., Gonzalez, G., Walker, K.M.: 2003, AJ 125, 2664
\bibitem{lineweaver}Lineweaver, C.H., Grether, D.: 2003, ApJ 598, 1350
\bibitem{marsakov}Marsakov, V.A., Shevelev, Yu.G.: 1995, BICDS 47, 13
\bibitem{naef}Naef, D., Mayor, M., Korzennik, S. G.: 2003, A\&A 410, 1051
\bibitem{neuhäuser}Neuh\"auser, R., Brandner, W., Eckart, A.: 2000, A\&A 354, 9
\bibitem{ng}Ng, Y.K., Bertelli, G.: 1998, A\&A 329, 943
\bibitem{patience}Patience, J., White, R.J., Ghez, A.M.: 2002, ApJ 581, 654
\bibitem{perryman}Perryman, M., Lindegren, L., Kovalevsky, J.: 1997, A\&A 323, 49
\bibitem{santos}Santos, N.C., Israelian, G., Mayor, M.: 2004, A\&A 415, 1153
\bibitem{udry}Udry, S., Mayor, M.: 2002, Springer, ISBN 3-540-42101-7, p.25-46
\bibitem{wilson}Wilson, J.C., Kirkpatrick, J.D., Gizis, J.E.: 2001, AJ 122, 1989
\bibitem{zucker}Zucker, S., Mazeh, T.: 2002, ApJ 568, 113
\end{thebibliography}
\end{document}